\documentclass[aps,prb,letterpaper,amsmath,amssymb,superscriptaddress,nofootinbib,longbibliography,twocolumn]{revtex4}
\usepackage{mathrsfs}
\usepackage{epsfig}
\usepackage{graphicx}
\usepackage{amsfonts}
\usepackage[figuresright]{rotating}
\usepackage{amssymb}
\usepackage{amsmath}
\usepackage{dcolumn}
\usepackage{bm}
\usepackage{color}
\usepackage[colorlinks=true,linkcolor=blue,citecolor=blue,anchorcolor=green,urlcolor=blue]{hyperref}
\usepackage{placeins}

\def\be{\begin{equation}}
\def\ee{\end{equation}}
\def\bea{\begin{eqnarray}}
\def\eea{\end{eqnarray}}

\def\ba{\begin{aligned}}
\def\ea{\end{aligned}}

\newcommand{\WQC} {Wilczek Quantum Center and Key Laboratory of Artificial Structures and Quantum Control, School of Physics and Astronomy, Shanghai Jiao Tong University, Shanghai 200240, China}

\newcommand{\Peking}{School of Physics, Peking University, 100871 Beijing, China}

\begin{document}
\title{Anomalous diffusion in quantum system driven by heavy-tailed stochastic processes}

\author{Chenyue Guo}
\email{guochenyue@sjtu.edu.cn}
\affiliation{\WQC}

\author{Yuchen Bi}
\affiliation{\Peking}


\begin{abstract}
In this paper, we study a stochastically driven non-equilibrium quantum system where the driving protocols consist of hopping and waiting processes. The waiting times between two hopping processes satisfy a heavy-tailed distribution. By calculating the squared width of the wavepackets, our findings demonstrate the emergence of various anomalous transport phenomena when the system remains unchanged within the heavy-tailed regime, including superdiffusive, subdiffusive, and standard diffusive motion. Only subdiffusion occurs when the system has evolved during the waiting process. All these transport behaviors are accompanied by a breakdown of ergodicity, highlighting the complex dynamics induced by the stochastic driving mechanism.

\end{abstract}
\maketitle

\section{introduction}
\label{sec1}
The continuous-time random walk (CTRW) model~[\onlinecite{LF_Montroll_1965, LF_GHW_1994, LF_KJ_SIM_2011}] is a mathematical framework used to describe the dynamic behavior of systems, particularly the diffusion process of particles in complex systems and non-uniform media~[\onlinecite{complex_S1975, complex_JP1990, complex_JK2000, complex_WM2013}].
This model applies to multiple fields such as physics~[\onlinecite{apply_phy_1974, apply_phy_1977, apply_phy_1985, apply_phy_1993, apply_phy_1997}], finance~[\onlinecite{apply_financial_1963, apply_financial_2003, apply_financial_2006}], and many others~[\onlinecite{apply_2002, apply_2003, apply_2013}], because it can handle discontinuities in time and space as well as long-range dependencies.
The CTRW model incorporates two independent stochastic  variables~[\onlinecite{complex_S1975, LF_Rep339}]: jump length $x$ and waiting time $t$, as described by the probability distribution functions (PDFs) $\lambda(x)$ and $\psi(t)$, respectively. 
They characterize a particle's waiting time $t$ at its initial position, then followed by a jump of length $x$, after which the process restarts.
When the PDFs $\lambda(x)$ and $\psi(t)$ exhibit finite first and second moments, such as $\lambda(x)$ and $\psi(t)$ follow a Poissonian and Gaussian distribution, the motion of system corresponds to normal diffusion~[\onlinecite{LF_KJ_SIM_2011}].
Conversely, if the waiting time PDF $\psi(t)$ adheres to a power-law distribution with a long tail, such that $\psi(t)\propto 1/t^{(1+\alpha)}$ with $0<\alpha<1$, the first moment, $\langle t \rangle=\int_0^{\infty} t \psi(t)dt$, diverges. The ensemble-averaged mean squared displacements satisfy $\langle x^2(t)\rangle_{ens}\sim t^{\alpha}$, indicating a subdiffusive motion~[\onlinecite{LF_Rep339, sub_2002}]. 
Additionally, the ensemble-averaged mean squared displacements do not equal the time-averaged mean squared displacements, indicating ergodicity breaking~[\onlinecite{Nonergodicity_1992, Nonergodicity_BE_2004, Nonergodicity_MG_2005, Nonergodicity_LA_2008, Nonergodicity_HY_2008, Nonergodicity_2023}].

A profusion of equilibrium quantum systems, including generic one-dimensional integrable models ~[\onlinecite{quantum_Zotos1997, quantum_book2005, quantum_Miller2020}], chaotic  systems~[\onlinecite{diffusion_Karrasch2014, diffusion_sub_2016, diffusion_Blake2017, diffusion_Friedman2020}] or 
disordered models~[\onlinecite{quantum_1958, Anomalous_Agarwal2015, Anomalous_Bar2015, MBL_Luitz2016}] typically exhibit ballistic, diffusive or subdiffusive (localization) motion.
The dynamics of non-equilibrium have garnered significant interest due to substantial experimental progress~[\onlinecite{exper_2008, exper_20182}].
By employing methods such as quenching~[\onlinecite{quantum_quench_2011, quantum_quench_2012}], driving~[\onlinecite{quantum_driven_2012, quantum_driven_cai, quantum_driven_2020, quantum_driven_guo}] or coupling to environment~[\onlinecite{quantum_coup_2013, quantum_coup_20132, quantum_coup_2016, quantum_coup_2022, quantum_coup_2023}], 
one can drive the system far from equilibrium and then analyze the transport phenomena emerged in the non-equilibrium system. 
Notably, recent studies~[\onlinecite{super_Wang2023,super_Wang2024}] have unveiled an innovative approach that demonstrates the emergence of superdiffusion by coupling quantum lattice systems with multi-site dephasing dissipation. 
Ref~[\onlinecite{super_Wang2023}] provided an insightful theoretical arguement using the framework of L$\rm{\acute{e}}$vy walk.
The L$\rm{\acute{e}}$vy walk (L$\rm{\acute{e}}$vy flight) model, with a finite (infinite) velocity of a random walker, is known for generating anomalous diffusion due to jump determined by the heavy-tailed distribution~[\onlinecite{review_2015}].
And this model is often regarded as the theoretical background for explaining superdiffusion~[\onlinecite{super_Wang2023,super_Wang2024, review_2015, levy_Zheng_2018, levy_2020, levy_2022}].

In this paper, we study a stochastically driven non-equilibrium quantum system. 
Unlike previous studies on stochastically driven systems~[\onlinecite{Stochastic_2017, Stochastic_2018, Stochastic_2020, Stochastic_2021, Stochastic_2022}], the driving time between two hopping processes satisfies the power-law PDF, $\psi(t)\propto 1/t^{(1+\alpha)}$ with $0<\alpha<1$, a typical heavy-tailed distribution.
It is noteworthy that the case is the quantum analog of the CTRW model. The toy model is composed of two parts, one depicted by Hamiltonian. (\ref{3}), which exhibits ballistic transport behavior distinct from the diffusion seen in classical random walks (Brownian motion) due to phase coherence and quantum interference effects. We are particularly interested in the transport behavior of the quantum system after introducing an evolution time $t$ distributed as a heavy-tailed distribution. This process is governed by another Hamiltonian. (\ref{2}). Here, we use the width of the wavepacket to characterize the system's dynamics.
For the case $V=0$ in Hamiltonian. (\ref{2}), we observe that the ensemble-averaged squared width of the wavepackets grows with time as $t^{2\alpha}$, corresponding to subdiffusive, superdiffusive and normal diffusive motion, with $\alpha$ varying in different intervals.
Furthermore, we calculate both the time-averaged squared width of the wavepackets and its ensemble average. It grows with time as $t^{1+\alpha}$. These two types of averages do not converge, leading to nonergodicity. 
For the case where $V\neq 0$, we observe that the ensemble-averaged 
(time-averaged) squared width of the wavepackets grows with time as $t^{\alpha}$ ($t^{0.75+\alpha}$), corresponding to subdiffusive (diffusive) motion, indicating the nonergodicity reoccurs.

The structure of the paper is organized as follows: Sec. \ref{sec2} presents the single-particle models defined in a one-dimensional (1D) lattice and methods used in this paper. In Sec. \ref{sec3}, we explore the transport behavior of ensemble-averaged and time-averaged squared width of the wavepackets. Sec. \ref{sec4} provides the conclusion and outlook.

\begin{figure}
\centering
 \includegraphics[width=8.0cm]{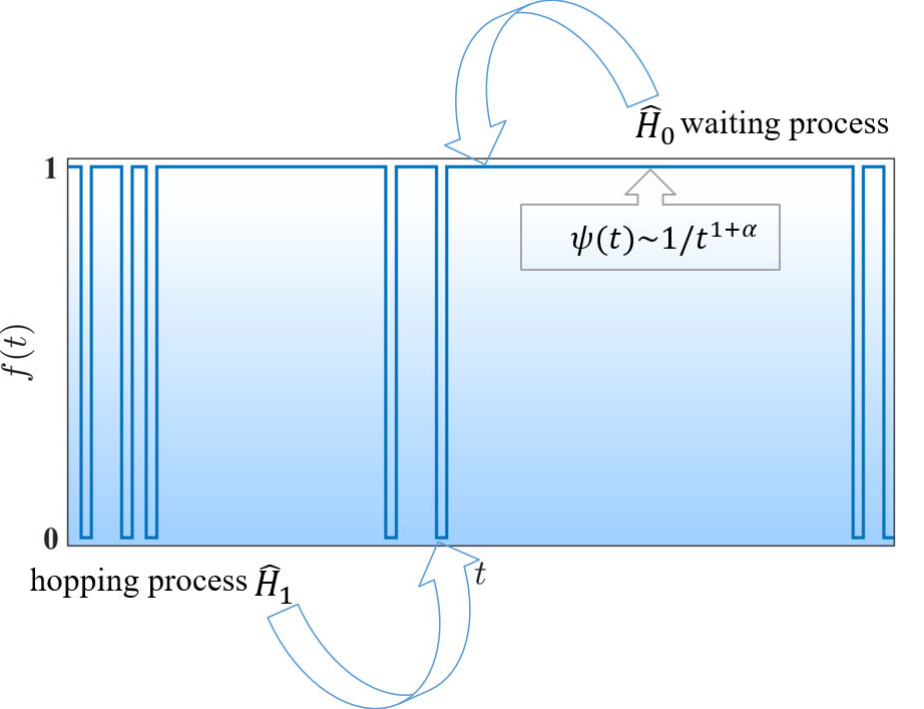}
 \caption{(Color online). Example of a typical trajectory of the stochastic driven quantum system. Control parameter $f(t)$ in the Hamiltonian. (\ref{1}) equals to $1$ or $0$ corresponds to the duration of $\hat{H}_{0}$ and $\hat{H}_{1}$, respectively. The time for waiting process follows the power law distribution.}
 \label{fig.0}
\end{figure}

\begin{figure}
\centering
 \includegraphics[width=8.5cm]{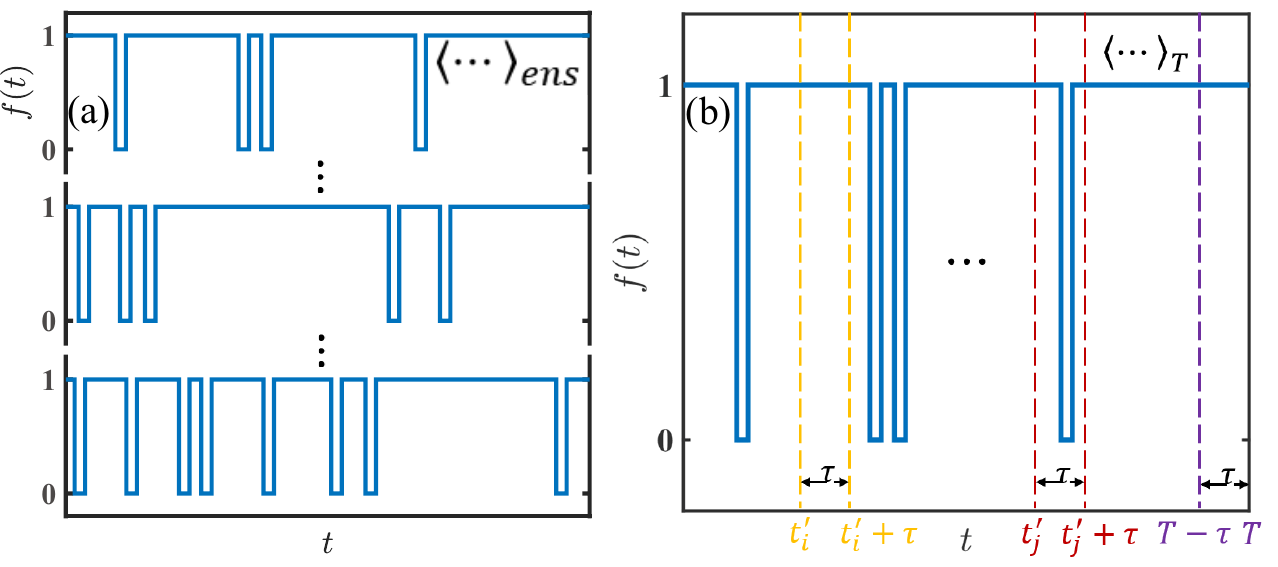}
 \caption{(Color online). Illustration of ensemble-averaged and time-averaged method in this paper. (a) $\langle \cdots\rangle_{ens}$ represents the quantity over different waiting time PDF trajectories unless otherwise stated. For $t\in[0,T]$, each distribution of $f(t)$ represents a realization of the PDF $\psi(t)$. (b) $\langle \cdots \rangle_{T}$ represents the quantity over different time segments within a single realization of the waiting time PDF trajectory.}
 \label{fig.02}
\end{figure}

\begin{figure*}
\centering
 \includegraphics[width=17.8cm]{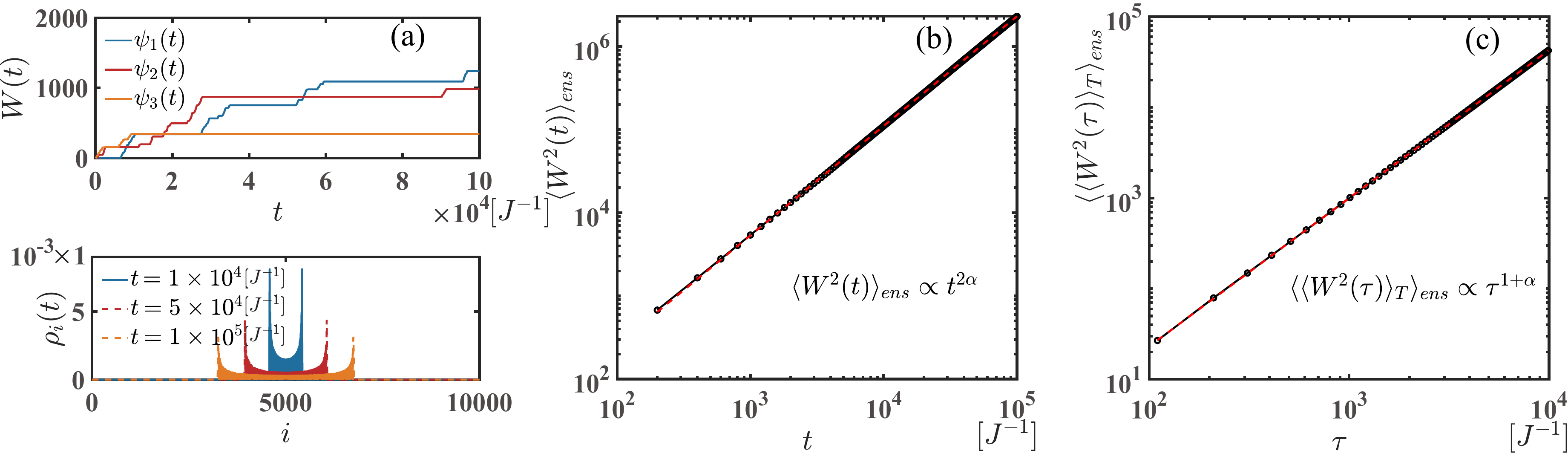}
 \caption{(Color online). (a) Width of the wavepacket under three different waiting-time PDFs (upper panel). Lower panel: density distributions of the wavepacket at different times for a single realization of $\psi(t)$. (b) Ensemble-averaged squared width of the wavepackets. (c) Ensemble time-averaged squared width of the wavepackets. The solid (dashed) black (red) line with circular markers represents the real (fitted) data. The system size $L=10000$. Other parameters $\Delta t=0.1J^{-1}$, $V=0$ and $\psi(t)\propto 1/t^{(1+\alpha)}$ with $\alpha=0.65$.}
 \label{fig.1}
\end{figure*}

\section{Model and method}
\label{sec2}
We consider a 1D single-particle model with Hamiltonian defined as:
\be 
\label{1}
\hat{H}(t)=f(t)\hat{H}_0+[1-f(t)]\hat{H}_1,
\ee
where
\begin{align}
\label{2}
\hat{H}_0&=\sum_{i}{(-1)^iV\hat{c}_i^\dagger \hat{c}_{i}},\\
\label{3}
\hat{H}_1&=\sum_{i}{-J\hat{c}_i^\dagger \hat{c}_{i+1}+h.c.}.
\end{align}
Where $\hat{H}_1$ and $\hat{H}_0$, including the free fermion hopping term and on-site potential term, correspond to the process of hopping and waiting governed by $f(t)$, respectively.
$\hat{c}_i$ ($\hat{c}_i^\dagger$) denotes the annihilation (creation) operator of a spinless fermion on site $i$. The parameter $J$ represents the amplitude for nearest-neighbor (NN) single-particle hopping and $V$ is the on-site chemical potential. Open boundary condition is adapted.

Starting from an initial state $\phi(t=0)=\delta(i=L/2)$ where a particle is located at site $i=L/2$, the wave function $\phi(t)$ evolves according to the Schr$\ddot{\rm{o}}$dinger eqation $i\hbar\partial \phi(t)/\partial t=\hat{H}_0(t)\phi(t)$, with the time $t$ for the waiting process following a power-law distribution $\psi(t)\propto1/t^{1+\alpha}$, with $\alpha=0.65$ in the main text [it is applicable for other $\alpha$ values shown in Appendix \ref{secs1}]. 
Then the particle starts hopping governed by the Hamiltonian $\hat{H}_1$ with $i\hbar\partial \hat{H}_1(t)/\partial t=\hat{H}_1(t)\phi(t)$, and the process is renewed. 
For simplification, the hopping time in the quantum model is quantified as lasting $t=1J^{-1}$.

To quantify the transport behavior emerged in the driven quantum model, we introduce the squared width of the wavepacket $W^2(t)$, defined as:
\be 
\label{4}
W^2(t)=\sum_i\rho_i(t)[i-\bar{x}(t)]^2,
\ee
where the density distribution $\rho_i(t)=\langle \phi(t)|\hat{c}_i^\dagger \hat{c}_{i}|\phi(t)\rangle$ and the center of matter (COM) of the wave packet defined as $\bar{x}(t)=\sum_ii\rho_i(t)$.
We can calculate the ensemble-averaged squared width of the wavepacket $\langle W^2(t)\rangle_{ens}$ over different waiting time PDF trajectories, as illustrated in Fig.~\ref{fig.02}\hyperref[fig.02]{(a)}.

In order to calculate the time-averaged squared width of the wavepackets, $\langle W^2(\tau)\rangle_{T}$, for a given waiting time PDF trajectory where $T$ is fixed,
we consider the time evolution of the wave function over a time $\tau$ at different initial positions $t_{i}'$, as illustrated in Fig.~\ref{fig.02}\hyperref[fig.02]{(b)}. Specifically, the time evolution of the wave function is expressed as:
$|\phi(t_{i}'+\tau)\rangle=\mathcal{\hat{T}}exp[-i\int_{t_{i}'}^{\tau+t_{i}'}\hat{H}(t^{\prime\prime})dt^{\prime\prime}]|\phi(t_{i}')\rangle$.
Here, $\mathcal{\hat{T}}$ is the time-ordering operator and $\hat{H}(t^{\prime\prime})$ equals either $\hat{H}_{0}$ or $\hat{H}_{1}$ depending on whether $t^{\prime\prime}$ lies in the waiting or hopping process.  
For each finite total evolution time $T$, $t_{i}'\in[0,T-\tau]$ with $i=1,2,..,\frac{T-\tau}{dt_{i}'}$ and $T\gg\tau$, ensuring sufficient average samples for large values of $\tau$.
Calculating a series of squared width of the wavepackets defined in eq. (\ref{4}) for different $t_{i}'$, the time-averaged squared width of the wavepackets $\langle W^2(\tau)\rangle_{T}$ is defined as:
\be 
\label{5}
\langle W^2(\tau)\rangle_T=\frac{1}{T-\tau}\int_0^{T-\tau}[W(t_{i}'+\tau)-W(t_{i}')]^2dt_{i}',
\ee

Moreover, for each finite total evolution time $T$, this quantity is treated as a random variable. Therefore, we consider the ensemble average $\langle \langle W^2(\tau) \rangle_{T}\rangle_{ens}$ over different waiting time PDF trajectories. 
In the main text, the magnitude of $T$ is set to $\mathcal{O}(10)\tau$, which is sufficiently large, as explained in Appendix \ref{secs21}. The transportation exponents remain consistent across different values of $T$, confirming the adequacy of this choice for $T$.

\begin{figure*}
\centering
 \includegraphics[width=17.5cm]{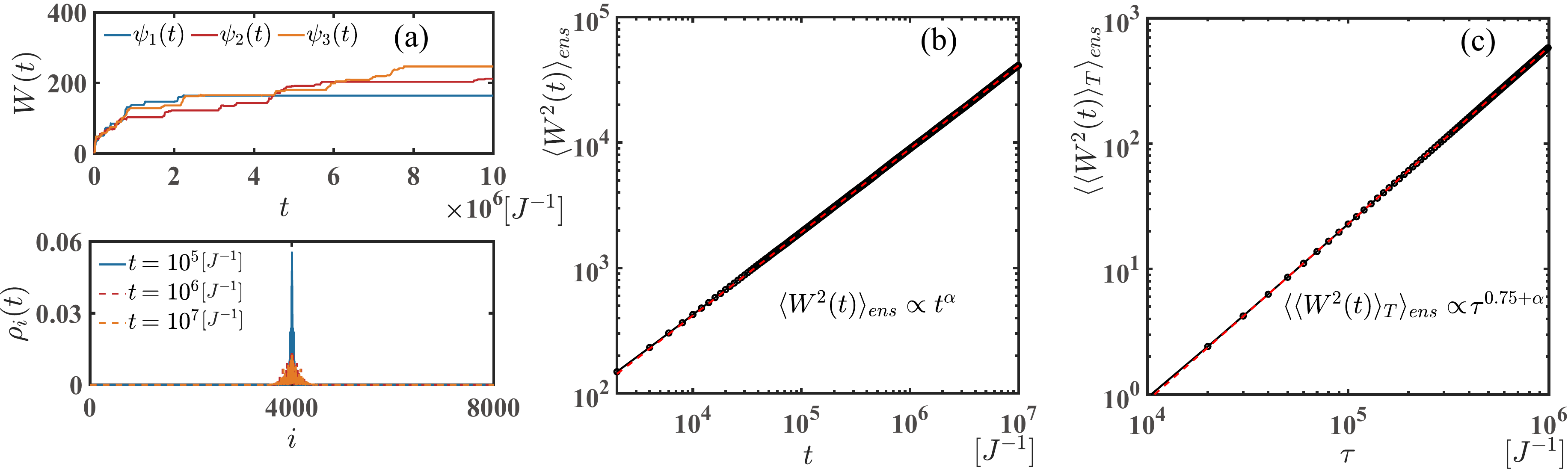}
 \caption{(Color online). (a) Width of the wavepacket under three different waiting-time PDFs (upper panel). Lower panel: density distributions of the wavepacket at different times for a single realization of $\psi(t)$. (b) Ensemble-averaged squared width of the wavepackets. (c) Ensemble time-averaged squared width of the wavepackets. The solid (dashed) black (red) line with circular markers represents the real (fitted) data. The system size $L=8000$. Other parameters $\Delta t=0.1J^{-1}$, $V=4J$ and $\psi(t)\propto 1/t^{(1+\alpha)}$ with $\alpha=0.65$.}
 \label{fig.2}
\end{figure*}

\section{Results}
\label{sec3}
\subsection{The case for free fermion}
\label{sec31}
In this section, we first consider the wave function remaining unchanged during the waiting process, i.e., $V$ is always set to zero ($\textit{zero-potential}$).
Specifically, we start from an initial state $\phi(t=0)=\delta(i=L/2)$, the wave function $\phi(t)$ keeps unchanged [$\hat{H}_{0}= 0$] for waiting time $t$ following a power-law distribution. Then the particle starts hopping governed by the Hamiltonian $\hat{H}_1$, and the process is renewed.
 
The upper panel of Fig.~\ref{fig.1}\hyperref[fig.1]{(a)} intuitively exhibits three trajectories of the squared width of the wavepackets.
We choose the discrete time step $\Delta t=0.1J^{-1}$ and system size $L=10000$, both shown to be sufficiently accurate in Appendix \ref{secs2}. 
Considering $W(t)$ as a random variable dependent on different waiting-time PDFs $\psi_{i}(t)$, we calculate the ensemble-averaged squared width of the wavepackets $\langle W^2(t)\rangle_{ens}$ over $5\times10^4$ waiting time PDF trajectories, as shown as a log-log plot in Fig.~\ref{fig.1}\hyperref[fig.1]{(b)}. We observe that it grows with time $t$ as a function of $t^{1.3179}$ ($\sim t^{2\alpha}$). Given that the jumping process corresponds to a free fermion hopping model within $\sqrt{W^2(t)}\sim t$ (as shown in eq. \eqref{app7} of Appendix \ref{secs3}), we only restrict the particle waiting for time $t_i$ based on the jumping process.
As a result, the density distribution of the wavepackets maintains the same shape as observed in ballistic transport, although with a reduced velocity of evolution, as illustrated in the lower panel of Fig.~\ref{fig.1}\hyperref[fig.1]{(a)} and Fig.~\ref{fig.1}\hyperref[fig.1]{(b)}. In other words, the involvement of the heavy-tailed time distribution only affects the velocity of the wavepacket's spreading while maintaining the overall wave shape. 
Generally speaking, the occurrence of anomalous diffusive transport arises solely from the heavy-tailed distribution of the waiting time $t$, which shares the same origin as the continuous-time random walk (CTRW) mode. The CTRW model introduces a waiting time distribution between consecutive jumps. Therefore, the case of free fermions can be regarded as a quantum analog of $\lambda(x)=\delta(x-1)$ in the CTRW model to a certain extent. Additionally, we present a direct analytical derivation of the anomalous diffusion, $\langle W^2(t)\rangle_{ens}\sim t^{2\alpha}$, as detailed in Appendix \ref{secs3}.

On a parallel front, we calculate the time-averaged squared width of the wavepacket $\langle W^2(\tau) \rangle_{T}$ and its ensemble average $\langle \langle W^2(\tau) \rangle_{T}\rangle_{ens}$ over $5\times10^4$ waiting time PDF trajectories, as shown in Fig.~\ref{fig.1}\hyperref[fig.1]{(c)}. 
We choose the integral time interval $dt_{i}'=5J^{-1}$, as shown to be sufficiently accurate in Appendix \ref{secs22}. 
We observe the quantity $\langle \langle W^2(\tau) \rangle_{T}\rangle_{ens}$ grows with time as a function of $\tau^{1.6337}$ $(\sim \tau^{1+\alpha})$, is not consistent with $\langle W^2(t)\rangle_{ens}$, indicating nonergodicity. 
So far, providing an analytical proof of time averaging has been challenging because the wave function evolves according to the schr\"{o}dinger equation, the width of the wave packet at each time $t_i'$ is not an independent variable.

\subsection{The effect of potential}
\label{sec32}

In this section, we consider the effect of the on-site potential $V$. For simplicity, we set $V$ as a time-independent constant ($\textit{constant-potential}$).
Specifically, we start from an initial state $\phi(t=0)=\delta(i=L/2)$, the wave function $\phi(t)$ evolves according to $i\hbar\partial \phi(t)/\partial t=\hat{H}_0(t)\phi(t)$ for waiting time $t$ following a power-law distribution. Then the particle starts hopping governed by the Hamiltonian $\hat{H}_1$ with $i\hbar\partial \phi(t)/\partial t=\hat{H}_1(t)\phi(t)$, and the process is renewed.

\begin{figure}
\centering
 \includegraphics[width=7.5cm]{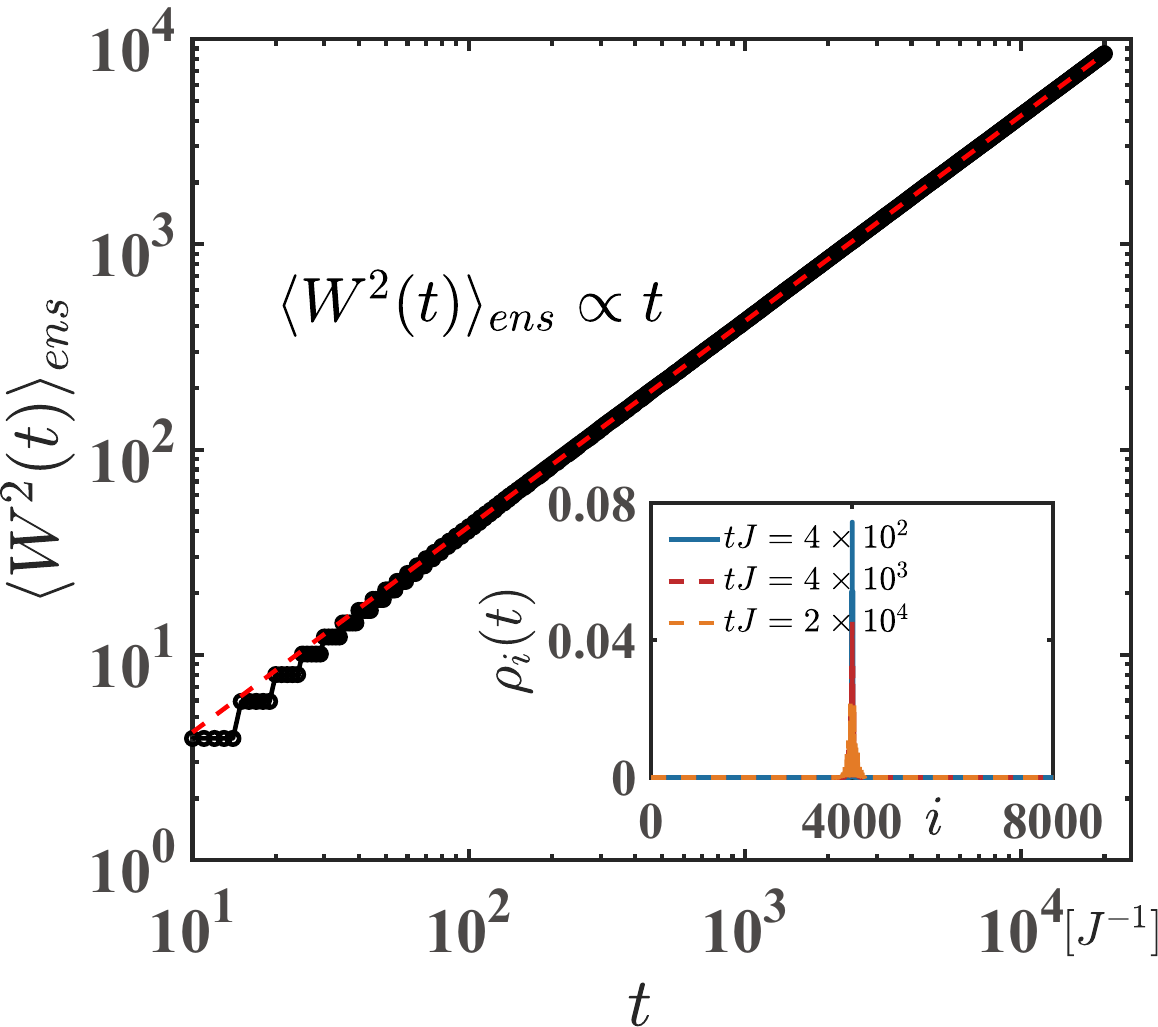}
 \caption{(Color online). Ensemble-averaged squared width of the wavepackets over $10^4$ realizations of $V$ PDF trajectories. The solid (dashed) black (red) line with circular markers represents the real (fitted) data. Inset: density distributions of the wavepacket at different times for a single realization of $\psi(V)$. The parameter $L=8000$, $\Delta t=0.01J^{-1}$ and $V(t)\propto 1/V^{(1+\alpha)}$ with $\alpha=0.65$.}
 \label{fig.3}
\end{figure}

The upper panel of Fig.~\ref{fig.2}\hyperref[fig.2]{(a)} illustrates three particle trajectories with the potential amplitude $V$ set to $4J$. The ensemble-averaged squared width of the wavepackets $\langle W(t)^2\rangle_{ens}$ is shown as a log-log plot in Fig.~\ref{fig.2}\hyperref[fig.2]{(b)}. 
We observe that it grows with time $t$ as a function of $t^{0.6650}$ ($\sim t^{\alpha}$) which is different from the case mentioned in Sec. \ref{sec31}. In this case, the unitary time evolution operator during the waiting process is given by $e^{-i\hat{H}_{0}t}=e^{\pm iVt}$ [$H_{0}$ is diagonal], where the absolute value of global phase factor $Vt$ can become significantly large due to the PDF $\psi(t)$ satisfied a power law distribution.
To determine which key element, $V$ or $t$, is more significant, we make $V$ to be time-dependent, following the same power law distribution $\psi(V)\propto 1/V^{(1+\alpha)}$, while setting the waiting time to be constant $t=4J^{-1}$ to ensure the same global phase factor 
[$e^{\pm i4\psi(t)} \sim e^{\pm i\psi(V)4}$]. The results are shown in Fig.~\ref{fig.3}. It reveals that the ensemble-averaged width of the wavepackets $\langle W^2(t)\rangle_{ens}\sim t$ and $\langle W(t)\rangle_{ens}\sim t^{0.5}$ [normal diffusion], indicating that the chemical potential $V$, with significant changes, is treated as white noise. 
Moreover, the lower panel of Fig.~\ref{fig.2}\hyperref[fig.2]{(a)} demonstrates that the density distribution of the wavepackets approximates a Gaussian distribution, corresponding to anomalous diffusive transport behavior ($\sqrt{\langle W(t)^2\rangle_{ens}}\sim t^{\alpha/2}$ with $\alpha<1$ shown in Fig.~\ref{fig.2}\hyperref[fig.2]{(b)}) due to the shifted velocity of evolution. 
Briefly speaking, the presence of anomalous diffusive transport shown in $\langle W^2(t)\rangle_{ens}$ is a consequence of the heavy-tailed distribution of $t$ for the waiting process, which shift the velocity of the wavepacket spreading.

On the other hand, we calculate the time-averaged squared width of the wavepacket $\langle W^2(\tau) \rangle_{T}$ and its ensemble average $\langle \langle W^2(\tau) \rangle_{T}\rangle_{ens}$, as shown in in Fig.~\ref{fig.2}\hyperref[fig.2]{(c)}. 
We choose the integral interval $dt_{i}'=50J^{-1}$, as shown to be sufficiently accurate in Appendix \ref{secs22}. 
The result reveals that the quantity grows with time as $\tau^{1.4126 }$ $(\sim \tau^{0.75+\alpha})$, which is not consistent with $\langle W^2(t)\rangle_{ens}$, indicating nonergodicity once again. 

\section{Conclusion and outlook}
\label{sec4}
In conclusion, our work explores the quantum analog of the CTRW model, revealing various anomalous diffusive transport behaviors accompanied by the breakdown of ergodicity.
For the \textit{zero-potential} case, we observe that the ensemble-averaged squared width of the wavepackets grow with time as $t^{2\alpha}$, encompassing subdiffusive, superdiffusive and the standard diffusive motion. Additionally, the time-averaged squared width of wavepackets and its ensemble average  grow with time as $t^{1+\alpha}$. 
The lack of convergence between these two averages indicates ergodicity breaking. 
%
For the \textit{constant-potential} case, the ensemble-averaged squared width of the wavepackets grow with time as $t^{\alpha}$, corresponding to subdiffusive motion. Meanwhile, $\langle \langle W^2(\tau) \rangle_{T}\rangle_{ens}$ grows with time as $\tau^{0.75+\alpha}$, signifying the recurrence of nonergodicity. 
The results highlight the abundant dynamics induced by the stochastic driving mechanism.

The power-law distribution is known for its ability to describe complex systems with a wide range of temporal scales.
Up to this point, we have somewhat artificially set the long-tailed waiting time in our models. This raises an intriguing question: can we develop a model in which the time correlation of physical quantities naturally conforms to a power-law probability distribution? If successful, such a model would likely exhibit richer dynamics compared to those governed by exponential decay interactions.

\section{Acknowledgments}
\label{sec5}
We thank Igor M. Sokolov for helpful and insightful suggestions on the continuous-time random walk. We also thank Zi Cai for the valuable discussions and guidance throughout this work, Hongzheng Zhao for useful discussions. This work is supported by the National Key Research and Development Program of China (Grant No. 2020YFA0309000), NSFC of China (Grant No.12174251), the Natural Science Foundation of Shanghai (Grant No.22ZR142830), and the Shanghai Municipal Science and Technology Major Project (Grant No.
2019SHZDZX01).

\begin{figure*}
\centering
 \includegraphics[width=17.0cm]{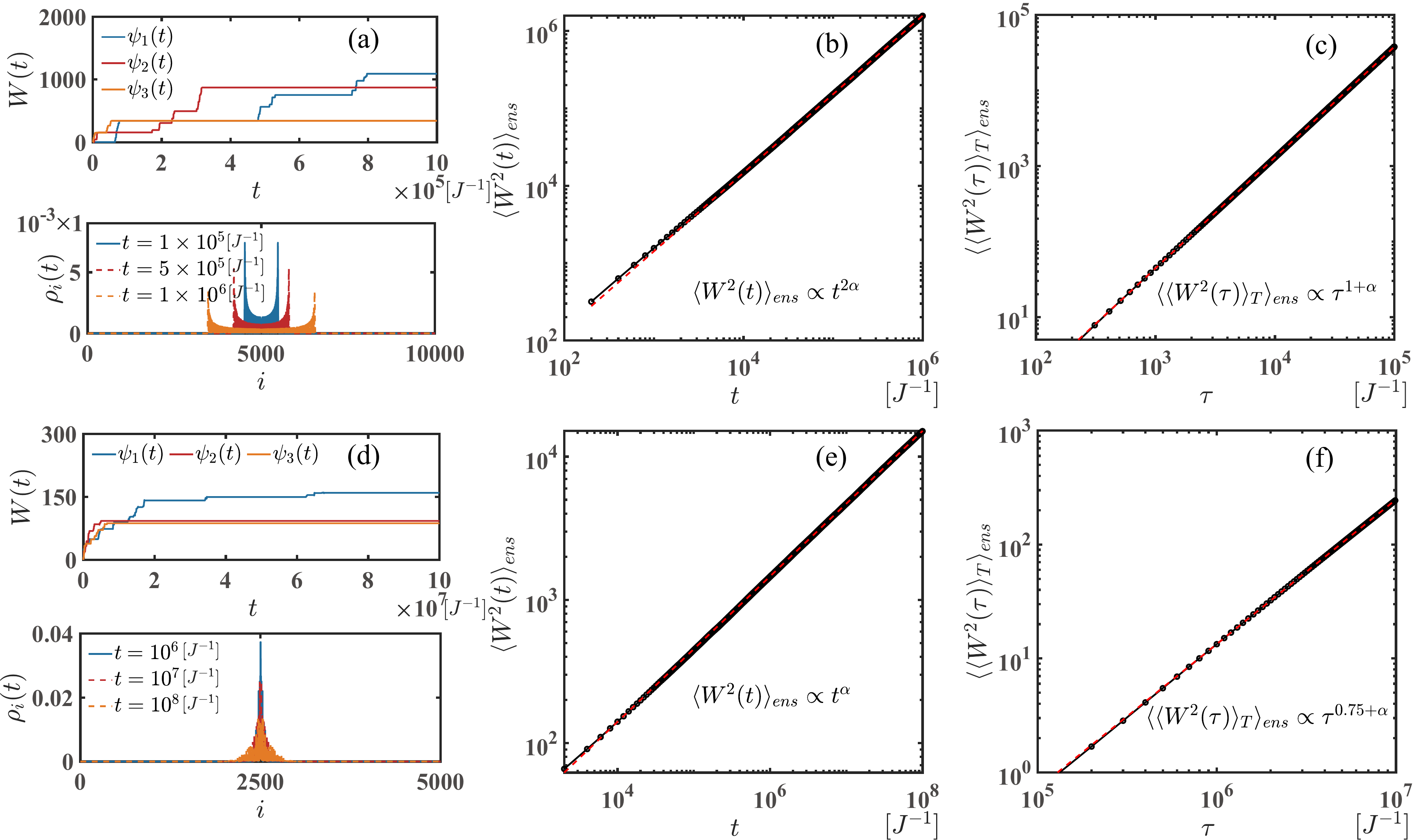}
 \caption{(Color online). The upper panel of (a)(d): Width of the wavepacket under three different waiting-time PDFs. The lower panel of (a)(d): density distributions of the wavepacket at different times for a single realization of $\psi(t)$. (b)(e) [(c)(f)] Ensemble-averaged [time-averaged] squared width of the wavepackets over $5\times10^4$ samples. The solid (dashed) black (red) line with circular markers represents the real (fitted) data for the right panel. The parameters $\Delta t=0.1J^{-1}$, $L=10000$, $V=0$ [$L=5000$, $V=4J$] for (a)-(c) [(d)-(f)] and $\alpha=0.5$.}
 \label{fig.4}
\end{figure*}

\begin{figure}
 \includegraphics[width=8.5cm]{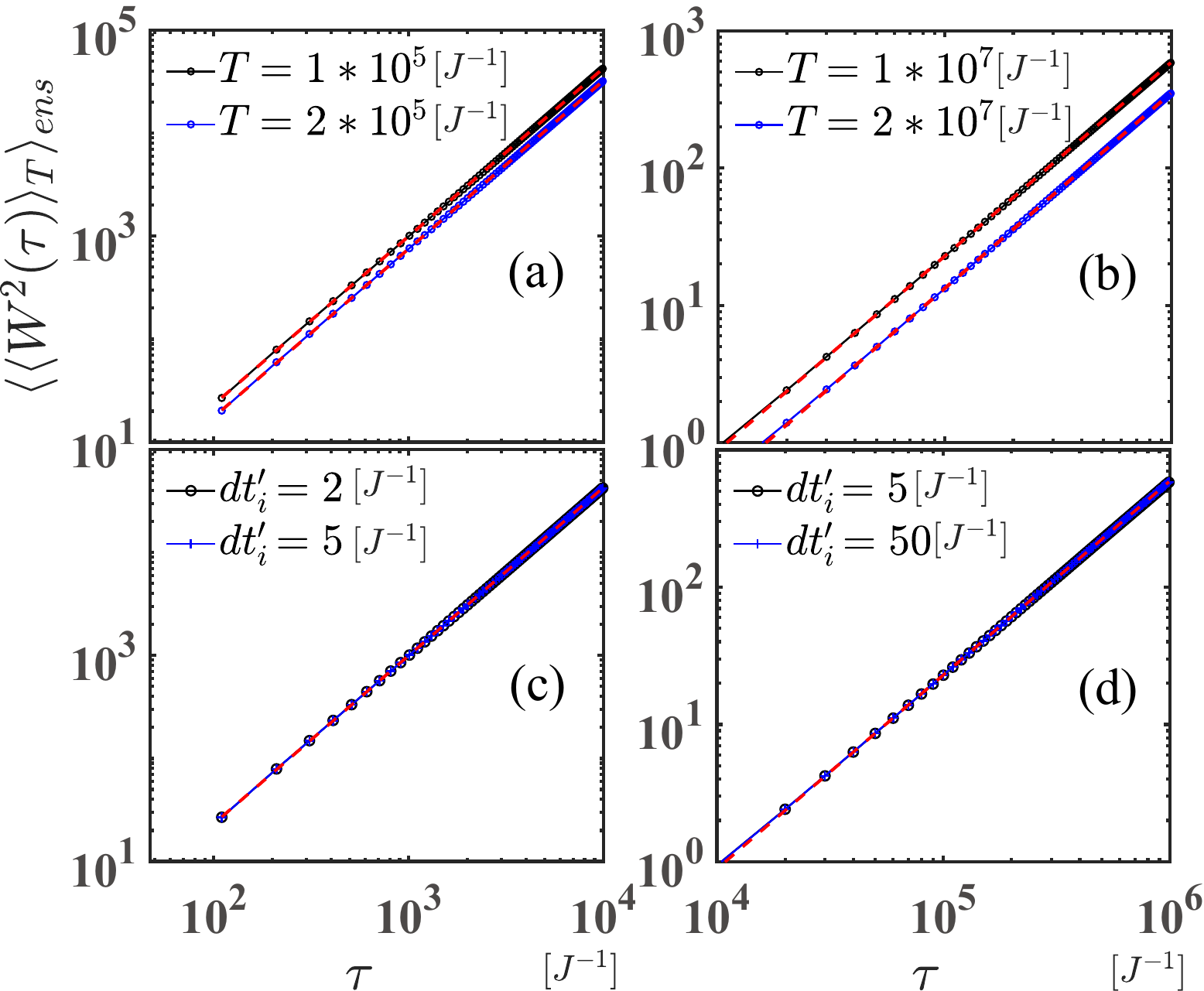}
 \caption{(Color online). Comparison between the Ensemble-averaged width of the wavepackets with different $T$ with $V=0$ (a) and $V=4J$ (b).  Comparison between the Ensemble-averaged width of the wavepackets with different integral time interval $dt_{i}'$ in eq. (\ref{5}) with $V=0$ (c) and $V=4J$ (d). The system size $L=10000$ for (a) and (c), $L=8000$ for (b) and (d). Other parameter $\alpha=0.65$.}
 \label{fig.5}
\end{figure}

\appendix

\section{universal of numerical results}
\label{secs1}
In the main text, we chose $\alpha=0.65$ and the results remain consistent for other values of $\alpha$ ranging between 0 and 1. As shown in the upper panel of Fig.~\ref{fig.4} for $\alpha=0.5$, it reveals that the ensemble-averaged squared width of the wavepackets $\langle W^2(t)\rangle_{ens} \sim t^{1.0148}$ ($\sim t^{2\alpha}$) and time-averaged squared width of the wavepacket $\langle \langle W^2(\tau) \rangle_{T}\rangle_{ens} \sim t^{1.4668}$ ($\sim t^{1+\alpha}$).
The lower panel of  Fig.~\ref{fig.4} shows that the ensemble and time averages scaling as $\langle W^2(t)\rangle_{ens} \sim t^{0.5079}$ ($\sim t^{\alpha}$)and $\langle \langle W^2(\tau) \rangle_{T}\rangle_{ens} \sim t^{1.2685}$ ($\sim t^{0.75+\alpha}$), respectively. 
The power exponents as fitted from the log-log plot follow the same rule with $\alpha=0.65$, 
indicating the universality of the anomalous transport behavior.

\section{Convergence of numerical results}
\label{secs2}

\subsection{Total time dependence}
\label{secs21}
In Sec. \ref{sec31}, we choose the total time $T=10^5\sim \mathcal{O}(10)\tau$. Here, we check its effectiveness comparing the larger $T=2*10^5$, as shown in Fig.~\ref{fig.5}\hyperref[fig.5]{(a)}. It reveals that the power exponents as fitted from the log-log plot keep consistent with absolute error $\mathcal{O}(10^{-4})$, signaling of the total time $T$ satisfies the condition $T\gg\tau$. 
Similarly, the chosen $T=10^7\sim \mathcal{O}(10)\tau$ in Sec. \ref{sec32} is also reasonable, as shown in Fig.~\ref{fig.5}\hyperref[fig.5]{(b)}.

\subsection{Integral interval $dt_{i}'$ dependence}
\label{secs22}
In Sec. \ref{sec31}, we choose integral interval $dt_{i}'=5J^{-1}$. To verify the convergence of our results with respect to $dt_{i}'$, we selected a smaller $dt_{i}'=2J^{-1}$ and compared the results shown in Fig.~\ref{fig.5}\hyperref[fig.5]{(c)}. It reveals that the simulation results are consistent with each other, indicating that the $dt_{i}'$ chosen is small enough to neglect numerical errors caused by the integral discretization.
The same applies for Sec. \ref{sec32}.

\begin{figure}
 \includegraphics[width=8.0cm]{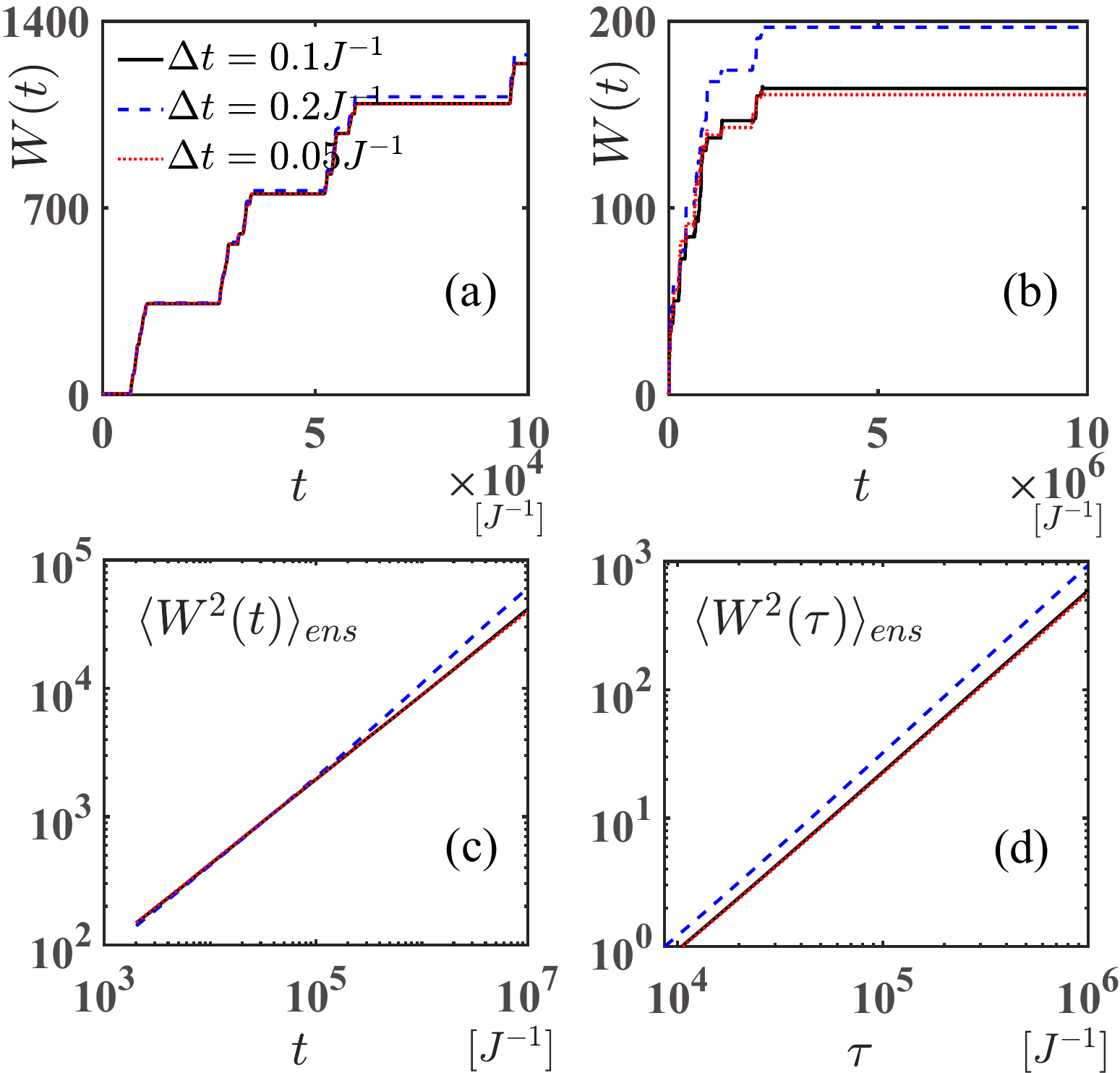}
 \caption{(Color online). Comparison between the width of the wavepacket for a waiting-time PDF with different $\Delta t$ with $V=0$, $L=10000$ (a) and $V=4J$, $L=8000$(b)-(d). (c) [(d)] Comparison between the ensemble-averaged [time-averaged] squared width of the wavepackets over $5\times10^4$ samples with different $\Delta t$. $\alpha=0.65$.}
 \label{fig.6}
\end{figure}

\begin{figure}[t]
 \includegraphics[width=8.0cm]{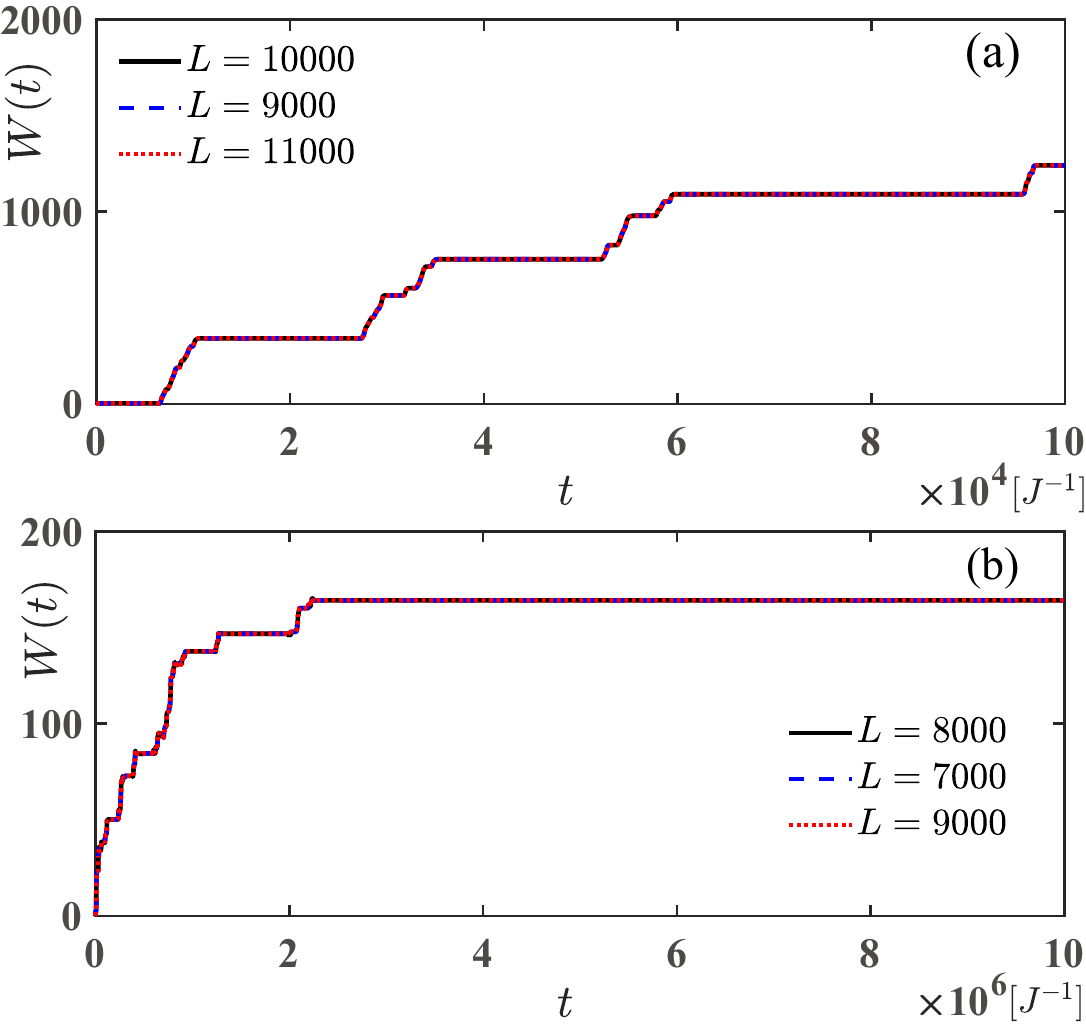}
 \caption{(Color online). Comparison between the width of the wavepacket for a waiting-time PDF with different system size $L$ with $V=0$ (a) and $V=4J$ (b). The parameter $\Delta t=0.1J^{-1}$ and $\alpha=0.65$.}
 \label{fig.7}
\end{figure}

\subsection{Discrete time step dependence}
\label{secs23}
Throughout the main text, we selected a discrete time step of $\Delta t = 0.1 J^{-1}$. The choice of $\Delta t$ is subtle because part of the system's evolution time is composed of waiting time, which directly depends on $\Delta t$. To verify the convergence of our results with respect to $\Delta t$, we selected different $\Delta t$ values of $0.2 J^{-1}$, $0.1 J^{-1}$, and $0.05 J^{-1}$, and compared their results. 
For the \textit{zero-potential} shown in Fig.~\ref{fig.6}\hyperref[fig.6]{(a)}, our simulation results are consistent with the smaller $\Delta t=0.05 J^{-1}$, indicating that the $\Delta t$ chosen is small enough to neglect numerical errors caused by the time discretization. 
For the \textit{constant-potential} shown in Fig.~\ref{fig.6}\hyperref[fig.6]{(b)}, the results differs for different $\Delta t$,  we infer that this is because the system continuously evolves due to the existence of $V$, so the state is slightly distinct for different $\Delta t$ when the hopping process starts, and the distinction can be amplified during time evolution. 
However, the non-convergence can be revised by the large number of samples because the small difference between these $\Delta t$ values almost does not affect the distribution of time $\psi(t)$. As shown in Fig.~\ref{fig.6}\hyperref[fig.6]{(c)-(d)}, the ensemble-averaged (time-averaged) squared width of wavepacket is converged to the smaller $\Delta t=0.05J^{-1}$, indicating that the $\Delta t$ chosen in our simulation is sufficient.

\subsection{System size dependence}
\label{secs24}
The main text outlines a single-particle simulation on a 1D lattice with finite system sizes of 
$L=10000$ for the \textit{zero-potential} case and $L=8000$ for the \textit{constant-potential} case, which limits the maximum simulation time. After this period, the wave packet will reflect off the boundaries of the $1D$ lattice under open boundary conditions. It is necessary to check the system boundary dependence of our results. As shown in Fig.~\ref{fig.7}\hyperref[fig.7]{(a)} and \hyperref[fig.7]{(b)}, the results converge well for different system sizes, indicating we have ruled out boundary effects.

\section{Analytical derivation of numerical results}
\label{secs3}
In Sec. \ref{sec31}, we determine that the ensemble-averaged squared width of the wave packets, $\langle W^2(t)\rangle_{ens}$, increases with time $t$ as $t^{1.3179}$ ($\sim t^{2\alpha}$), as illustrated in the log-log plot in Fig.~\ref{fig.1}\hyperref[fig.1]{(b)}. In this section, we provide a direct analytical derivation of the quantum evolution dynamics. In this approach, we first describe the transport dynamics governed exclusively by the hopping Hamiltonian $\hat{H}_1$. After introducing the waiting time, we calculate the mean number of hopping processes, $\langle n(t) \rangle$, that occur within time $t$, which determines the effective time $\langle n(t) \rangle \cdot t_h$. This effective time corresponds precisely to the evolution time of the wave function, which remains unchanged during the rest period $(\hat{H}_0=0)$. The details are outlined below.

\subsection{Ballistic transport governed by the hopping term}
\label{secs31}
By introducing the Fourier transformation $\hat{a}_j = \frac{1}{\sqrt{L}} \sum_k \hat{a}_k e^{ijk}$, Eq. (\ref{3}) can be rewritten as:   
\be 
\label{app1}
\hat{H}_1=\sum_k \epsilon_k \hat{a}_k^\dagger \hat{a}_k,
\ee
where $\epsilon_k=-2J \cos(k)$ represents the energy associated with the momentum $k$.
The initial state in position space, $\phi(j,0) = \delta_{j,\frac{L}{2}}$, can be transformed into momentum space as $\phi(k,0) = \frac{1}{\sqrt{L}}$. Then the wave function of momentum $k$ at time $t$ in momentum space is given by:
\begin{equation}
\label{app2}
\phi(k,t) = e^{-iH_k t} \phi(k, 0) = \frac{1}{\sqrt{L}} e^{i2Jt \cos(k)},
\end{equation}
Then, the wave function $\phi(j,t)$ in real space becomes:
\begin{align}
\label{app3}
\phi(j, t) &= \frac{1}{\sqrt{L}} \sum_k e^{ijk} e^{i2Jt\cos(k)} 
\xrightarrow{L \to \infty} \nonumber \\
&=\int_{-\pi}^\pi \frac{dk}{2\pi} e^{ijk} e^{i2Jt\cos(k)} \nonumber \\
&= i^j J_j(2Jt), 
\end{align}
where $J_j(z)=\int_{-\pi}^\pi \frac{dk}{2\pi} e^{ijk} e^{-iz\sin(k)}$ with $z=2Jt$ denotes the $j$-th Bessel function. This demonstrates that the density distribution of the wave function at position $j$, i.e., $|\phi(j, t)|^2$, is proportional to the Bessel function $J^2_j(2Jt)$.

The squared width of the wavepacket $W^2(t)$ defined as eq. (\ref{4}), $W^2(t)=\sum_j\rho_j(t)[j-\bar{x}(t)]^2$, where the lattice site index $i$ to $j$ to distinguish it from the imaginary number $i$. This simplifies to
\begin{equation}
\label{app4}
W^2(t) = \sum_j j^2 |\phi_j(t)|^2 = \sum_j j^2 J_j^2(2Jt),
\end{equation}
where $\bar{x}(t)=0$. This is because, in the initial state, a particle is located at the middle site $L/2$, labeled as $0$ in numerical simulations with $j\in [-L/2,L/2)$. And the center of mass (COM) of the wave packet remains at the middle lattice site  throughout the entire time evolution [as confirmed by numerical results shown in the lower panel of Fig.~\ref{fig.1}\hyperref[fig.1]{(a)} and Fig.~\ref{fig.2}\hyperref[fig.2]{(a)}], allowing the simplification $\bar{x}(t)=0$.

Next, applying partial derivatives with respect to $\theta$ to the generator function of Bessel function:
\begin{align}
\label{app5}
\frac{\partial}{\partial \theta} \left[ e^{i z \cos \theta} \right]=
\frac{\partial}{\partial \theta} \left[ \sum_j i^j J_j(z) e^{ij \theta} \right],
\end{align}
we obtain $-i z \sin \theta e^{i z \cos \theta} = \sum_j i^{j+1} j J_j(z) e^{ij \theta}$, and further simplifies to:
\begin{align}
\label{app6}
\sum_{j,j'} i^{j+1} i^{j'+1} j j' J_j(z) J_{j'}(z) \delta_{j,j'} &= \int_{-\pi}^{\pi} z^2 \sin^2\theta \frac{d\theta}{2\pi}  \nonumber \\
\xrightarrow{\text{simplify}} \sum_{j} j^2 J_j^2(z) &= \frac{z^2}{2}.  
\end{align}

Thus, eq. \eqref{app4} can be rewritten as:
\begin{align}
\label{app7}
W^2(t) = \sum_j j^2 J_j^2(2Jt) = 2J^2t^2, \quad \text{or} \nonumber \\
\sqrt{W^2(t)}=\sqrt{2}Jt \sim t^1.
\end{align}
So far, we have derived the typical transport relation for ballistic transport. 

\subsection{Anomalous transport after introducing the waiting process}
\label{secs32}
Next, we consider the impact of the waiting process. It is known that the squared width of the wavepackets, $W^2(t)\sim t^2$, so it is necessary to calculate the mean squared effective time, $\langle n^2(t) \rangle_{ens} \cdot t_h^2$.

Assume there are $n$ occurrences of hopping process within time $t$, each accompanied by a waiting time. 
The probability of $n$ hopping processes occurring within time $t$ is given by:
\begin{equation}
\label{app8}
P_n(t) = \int_0^{t-nt_h} \int_{t-\xi-nt_h}^{+\infty} \psi_n(\xi) d\xi \psi(\xi_{n+1}) d\xi_{n+1} 
\end{equation}

Here, $\xi$ represents the sum of the first $n$ waiting times, $t_w^{(1)} + t_w^{(2)} + \cdots + t_w^{(n)}$, satisfying $t_w^{(1)} + t_w^{(2)} + \cdots + t_w^{(n)} < t < t_w^{(1)} + t_w^{(2)} + \cdots + t_w^{(n+1)}$. 
$\xi_{n+1}$ represents the waiting time for the $(n+1)$-th jump, $t_w^{(n+1)}$.
In the main text, the time for a single hopping process is denoted as $t_h$, where $t_h = 1 J^{-1} \ll \langle t_w \rangle \to \infty$. Therefore, eq. (C8) can be simplified to
\begin{equation}
\label{app9}
P_n(t)=\int_0^t \psi_n(\xi) [1 - \Psi(t - \xi)] d\xi
\end{equation}
where $\Psi(t-\xi)$ is the cumulative distribution function (CDF) of the waiting time for the $(n+1)$-th jump. $1-\Psi(t-\xi)$ represents the waiting time for the $(n+1)$-th jump is greater than $t-\xi$, ensuring that only $n$ hopping processes occur within time $t$.

Then the Laplace transform of probability $P_n(t)$ can be further expressed as:
\begin{equation}
\label{app10}
P_n(u) = \mathcal{L}\{P_n(t)\} = \mathcal{L}\{\psi_n(t)\} \cdot \mathcal{L}\{1 - \Psi(t-\xi)\}.
\end{equation}

Considering the $t_w^{(i)}$ are the independent variables satisfying the heavy-tailed distribution, the Laplace transform of 
$\psi_n(t)$ is given by:
\begin{align}
\label{app11}
\mathcal{L}\{\psi_n(t)\} &= \langle e^{-u(t_w^{(1)} + t_w^{(2)} + \cdots + t_w^{(n)})}\rangle  \nonumber \\
&=\langle e^{-u t_w^{(1)}} e^{-u t_w^{(2)}} \cdots e^{-u t_w^{(n)}}\rangle \nonumber \\
&=\langle e^{-u t_w^{(1)}}\rangle \langle e^{-u t_w^{(2)}}\rangle \cdots \langle e^{-s t_w^{(n)}}\rangle \nonumber \\
&=\psi^n(u)
\end{align}
where \(\psi(u)\) is the Laplace transform of $\psi(t)$. Similarly, $\mathcal{L}\{1 - \Psi(t- \xi)\} = \mathcal{L}\{1\} - \mathcal{L}\{\Psi(t-\xi)\} = \mathcal{L}\{1\} - \mathcal{L}\{1 \cdot \psi(t)\} = \frac{1}{u} - \frac{\psi(u)}{u}$. Substituting these results into eq. \eqref{app9}, we obtain reads $P_n(u) = \psi^n(u) \cdot \frac{1 - \psi(u)}{u}$.
Now the mean squared number of hopping process occurrences, $\langle n^2(u) \rangle_{ens}$, is given by:
\begin{align}
\label{app12}
\langle n^2(u) \rangle_{ens}&= \sum_n n^2 P_n(u) \nonumber \\
&= \frac{1 - \psi(u)}{u} \sum_n n^2 \psi^n(u) \nonumber \\
&= \frac{1 - \psi(u)}{u} \left( \frac{\partial}{\partial \psi(u)} + \frac{\partial^2}{\partial \psi(u)^2} \right) \sum_n \psi^n(u) \nonumber \\
&= \frac{2}{u} \left( \frac{\psi(u)}{1 - \psi(u)} \right)^2 + \frac{1}{u} \frac{\psi(u)}{1 - \psi(u)}.
\end{align}

In the main text, the waiting time $\psi(t)\propto 1/t^{(1+\alpha)}$ with $0<\alpha<1$ and its Laplace transforms read as $\psi(u)\sim 1-u^\alpha$. Substituting this expression into eq. \eqref{app12} and applying the inverse Laplace transform, we obtain:
\begin{equation}
\label{app13}
\langle n^2(t) \rangle_{ens} = 1 - \frac{3 \, t^\alpha}{\Gamma(1 + \alpha)} + \frac{2 \, t^{2\alpha}}{\Gamma(1 + 2\alpha)} \xrightarrow{t \to \infty} \sim t^{2\alpha}.
\end{equation}

So the efficient time for the hopping process is $t_{\text{eff}}=\langle n(t) \rangle_{ens} \cdot t_h$, which corresponds to the evolution time of the wave function. Using eq. \eqref{app7}, the ensemble averaged squared width of the wavepacket $\langle W^2(t)\rangle_{ens} $ is determined by:
\begin{align}
\label{app14}
\langle W^2(t)\rangle_{ens} &= \langle W^2(t_\text{eff})\rangle_{ens} \nonumber \\
&= 2 J^2 t_\text{eff}^2= 2J^2 t_h^2 \langle n^2(t)\rangle_{ens} \sim t^{2\alpha}
\end{align}
where $t_h=1J^{-1}$ in the main text. Therefore, we present a rigorous proof to support the numerical results.

\FloatBarrier

\bibliography{main}

\FloatBarrier

\end{document}